\documentclass[prb,preprint]{revtex4}

\pdfoutput=1

\usepackage{graphicx}

\begin{document}

\title{Coulomb-oscillator explanation of striped STM images of superconductive copper oxides\medskip }

\date{March 25, 2013} \bigskip

\author{Manfred Bucher \\}
\affiliation{\text{\textnormal{Physics Department, California State University,}} \textnormal{Fresno,}
\textnormal{Fresno, California 93740-8031} \\}

\begin{abstract}

Asymmetric scanning tunneling microscopy (STM) of the $CuO_{2}$ plane of
$Ca_{2-x}Na_{x}CuO_{2}Cl_{2}$, $x = 0.125$, shows a square domain structure with edge length four times the compound's lattice constant ${a_0}$ (Cu-O-Cu distance).  The domain structure is a direct consequence of the $4{a_0} \times 4{a_0}$ superlattice formed by vertical $Na^{+}$ pairs (oriented parallel to the crystal's $c$ axis) that substitute $Ca^{2+}$ ions.  The surrounding $O^{2-}$ ions are displaced away from, and the $Cu^{2+}$ ions toward the $Na^{+}$ pairs.  Contrary to the fourfold symmetry of the $CuO_{2}$ plane, the stable displacement configuration has a twofold symmetry, dominated by large and, respectively, small displacement of opposite $O^{2-}$ ions being nearest neighbors to each vertical $Na^{+}$ pair.  The ion displacements give rise to sufficient squeeze of certain $O^{2-}$ ions that, by the Coulomb-oscillator model of superconductivity, prevents lateral overswing of their excited $3s$ electrons.  The axial $3s$ Coulomb oscillations are predominantly oriented in the directions of $O^{2-}$ ion displacements. The observed ladder pattern in the domains provides a direct imaging of the $3s$ Coulomb oscillators.  The ``sidepieces'' of the ladders correspond to long unidirectional pathways for $3s$ electrons in the $CuO_{2}$ plane.  They account for superconductivity.  The findings lend support to the validity of the Coulomb-oscillator model of superconductivity.

\end{abstract}

\maketitle


$Ca_{2-x}Na_{x}CuO_{2}Cl_{2}$ and $Bi_{2}Sr_{2}Ca_{1-x}Dy_{x}Cu_{2}O_{8+\delta}$, like many other copper oxide compounds, become superconductive only when doped, with onset in the range $x \approx 0.05 - 0.10$.  Asymmetric scanning tunneling microscopy (STM) of the crystals in the superconducting state has revealed a unidirectional electron pattern on the atomic scale.\cite{1}  Various interpretations for this pattern have been proposed.\cite{1,2,3,4,5}  Here it is argued that the pattern is a manifestation of superconductive Coulomb oscillators. 

  The Coulomb-oscillator model of superconductivity\cite{6} attributes non-resistive electron current exclusively to the motion of the atoms' outer $s$ electrons through the nuclei of neighbor atoms (axial Coulomb oscillation).  Superconductivity is destroyed if the electrons' accompanying lateral oscillation, perpendicular to the internuclear line, extends between neighbor atoms.  The third tenet of the Coulomb-oscillator model of superconductivity is the \emph{squeeze effect}, whereby a compression of the atoms that provide the relevant outer $s$ electrons causes a reduction of lateral oscillation width.  The difference $\Delta y$ between the lateral oscillation width, $B0B$, and the lateral confinement limit, $\underbar{B0B}$, is a proportional measure for the critical temperature of superconductivity, $\underbar{B0B} - B0B \equiv \Delta y \propto T_{c}$. 

  Experimentally, the imaging method used in Ref. 1 is based on tunneling asymmetry, employing the ratio of STM tip-sample current under negative and positive bias.  It provides an expression for the imbalance of electron extraction (negative bias) and electron injection (positive bias).  For the samples of $Ca_{2-x}Na_{x}CuO_{2}Cl_{2}$ and $Bi_{2}Sr_{2}Ca_{1-x}Dy_{x}Cu_{2}O_{8+\delta}$ it was found that electron extraction is strongly favored over electron injection.  The degree of imbalance is imaged by darkness coding.

The two compounds investigated in Ref. 1 have distinctly different crystal structure (see Figs. 2ab of Ref. 1) except for a common $CuO_{2}$ plane, depicted here in Fig. 1. It has long been believed that superconductivity occurs in that plane.  In undoped $Ca_{2}CuO_{2}Cl_{2}$, the $CuO_{2}$ plane is sandwiched by layers of $Ca^{2+}$ ions, residing above and beneath the centers of the Cu-O-Cu squares.  
Neither a free $Cu^{2+}$ ion, nor an $O^{2-}$ ion in an alkaline-earth oxide---free $O^{2-}$ ions are unstable---possesses an outer $s$ electron as required by the Coulomb-oscillator model of superconductivity.  However, in the copper oxides of high-$T_{c}$ superconductivity the strong interionic Coulomb forces exert a compression of the conducting $CuO_{2}$ layer as a whole as well as of the $O^{2-}$ ions within that layer by their neighbor $Cu^{2+}$ ions (or in some cases $Cu^{3+}$ ions).  In reaction, the crystal reduces short-range repulsion between the $Cu^{2+}$ and $O^{2-}$ ions by an electron reconfiguration in $O^{2-}$ from $2p^{6}$ to $2p_a^1p_b^1p_c^23{s^2}$, 
that is, by reducing the occupancy of the $2p_{a}$ and $2p_{b}$ orbitals that are aligned along the $O$-$Cu$ axes in favor of excited, isotropic $3s$ orbitals.  It is those $3s$ electrons of extremely squeezed $O^{2-}$ ions that form the Coulomb oscillators in the oxide superconductors.  In the $CuO_{2}$ planes of the compounds under consideration,
$Ca_{2-x}Na_{x}CuO_{2}Cl_{2}$ and $Bi_{2}Sr_{2}Ca_{1-x}Dy_{x}Cu_{2}O_{8+\delta}$, the $3s$ electrons participate in Coulomb oscillations both between nearest-neighbor $O^{2-}$ ions (diagonally to the crystal's $a$ and $b$ axes) and between next-nearest $O^{2-}$ neighbors (parallel to the $a$ or $b$ axis).  The accompanying \emph{lateral} oscillations of the $3s$ electrons between \emph{nearest} $O^{2-}$ neighbors \emph{always} overswing, not only in the undoped crystal but also in the doped compound, treated below.  Diffusing throughout the $CuO_{2}$ layer, those $3s$ electrons are ruled out by the Coulomb-oscillator model of superconductivity from being superconductive.  As will become clear shortly, that mode of $3s$ electrons is lost to asymmetric STM imaging, too.

The lateral oscillations of the $3s$ electrons between \emph{next-nearest} $O^{2-}$ neighbors also (but barely) overswing in the undoped crystal.\cite{7}  However, in the \emph{doped} crystal the \emph{extra} squeeze of $O^{2-}$ ions by displaced $Cu^{2+}$ neighbors sufficiently reduces the width of lateral oscillation that accompanies $3s$ Coulomb oscillation between those next-nearest $O^{2-}$ neighbors.  Such ion displacements are indicated by arrows in the present Fig. 1 for the $CuO_{2}$ plane of $Ca_{2-x}Na_{x}CuO_{2}Cl_{2}$, $x = 0.125$.\cite{8}  The displacement-induced, extra squeeze of the $O^{2-}$ ions accounts for the emergence of superconductivity at a doping ratio $x \approx 0.05 - 0.10$ and for a rise of critical temperature $T_{c}$ with increased doping up to $x \approx 0.15$.  Moreover, those $3s$ electrons are amenable to asymmetric STM imaging.

In the case of doped $Ca_{2-x}Na_{x}CuO_{2}Cl_{2}$ with $x = 0.125 = 1/8$, a $Na^{+}$ ion substitutes every ${16^{th}}$ $Ca^{2+}$ ion of an undoped crystal.  The accompanying reduction of valence electrons can be ragarded as holes, ${e^ +}$, among the $3s$ electrons from excited $O^{2-}$ ions, participating in their activities in the same proportion as the doping ratio $x$. Minimal binding energy of the crystal is achieved when the doped $Na^{+}$ ions in the sandwiching layers coordinate their positions to form vertical $Na^{+}$ \emph{pairs} (parallel to the crystal's $c$ axis). For $x = 0.125$, each vertical $Na^{+}$ pair replaces every ${16^{th}}$ vertical $Ca^{2+}$ pair.  Expressed in terms of the crystal's lattice constant (Cu-O-Cu distance) $a_{0} = 3.85$ {\AA}, this imposes on the $CuO_{2}$ plane a square superstructure of $4{a_0} \times 4{a_0}$ domains.  The square frame in the present Fig. 1 delimits such a domain.  The crosses at the sides of the frame each mark the position of a vertical $Na^{+}$ pair residing in the planes above and beneath.  Experimentally, such square domains of $4{a_0} \times 4{a_0}$ extension are discernible in the STM images in Fig. 3 of Ref. 1.

A trend well known from lattice defects in ionic crystals, nearby $O^{2-}$ ions are displaced \emph{away} from the vertical pairs of substitutional $Na^{+}$ ions.  Somewhat counterintuitive with regard to space availability in terms of ionic radii---$r(Ca^{2+}) = 0.99$ {\AA}, $r(Na^{+}) = 0.95$ {\AA}---the reason is that the accumulate long-range Coulomb forces dominate over interionic short-range repulsion.  For a simplified explanation, one can think that the outward displacement of neighboring $O^{2-}$ ions results from their lesser electrostatic attraction to $Na^{+}$ than to $Ca^{2+}$.  Likewise the inward displacement of nearby $Cu^{2+}$ ions is caused by their lesser electrostatic repulsion from $Na^{+}$ than from $Ca^{2+}$.  There are two equilibrium configurations of displaced $O^{2-}$ ions:  The configuration with fourfold symmetry, dominated by \emph{equal} outward displacements of the four $O^{2-}$ ions that are nearest neighbors to a $Na^{+}$ pair (cross mark in the present Fig. 1), is unstable.  Any slight disturbance causes a transition to the stable equilibrium configuration of twofold (mirror) symmetry and presumably lower energy.  It is dominated by \emph{unequal} displacements of the four closest $O^{2-}$ ions away from the $Na^{+}$ pair---large outward displacements of two opposite closest $O^{2-}$ neighbors (white arrows up and down in Fig. 1) and small outward displacements of the other two closest $O^{2-}$ neighbors (to the left and right in Fig. 1).

Before turning to the question in which way the extra squeeze of $O^{2-}$ ions from those displacements affects lateral and axial oscillation of $3s$ electrons, we need to consider the energy situation.  The spatial confinement of the squeezed $O^{2-}$ ions gives rise to an increase of their electron energy ($2p \to 3s$).  Those $3s$ electrons either diffuse in the $CuO_{2}$ plane or, if subject to \emph{extra} squeeze from neighbor-ion displacement, perform laterally confined Coulomb oscillations (squeeze effect\cite{6}) with \emph{extra} increased energy.  Accordingly it is easier to extract a $3s$ electron from an extra squeezed region in the sample to an STM tip under negative bias than to inject an external electron under positive bias---a simple explanation for the observed tunneling anisotropy.  The dark regions in the STM images in Figs. 3, 4 and 6 of Ref. 1 can therefore be identified as laterally confined Coulomb oscillations of $3s$ electrons of extra squeezed $O^{2-}$ ions, swinging through their home nucleus, through their neighbor nucleus, and axially across the interionic region.

More needs to be learned about the exact mechanism of the squeeze effect in an atomic layer, particularly its dependence on atomic (ionic) displacement.  But three features stand out for the $O^{2-}$ ions in columns 4, 3 and 2 (and symmetrically 4', 3' and 2') in the present Fig. 1.  First, the $O^{2-}$ ions of column 4 are much squeezed by their hard $Cu^{2+}$ neighbors in columns 3 and 3''.\cite{9}  This causes a laterally confined Coulomb oscillation of $3s$ electrons along column 4, depicted in Fig. 1 by straight-line hatching.  Second, the additional squeeze of displaced $O^{2-}$ ions in column 2, experienced mutually with the soft nearest-neighbor $O^{2-}$ ions of column 1 (participating squeeze effect\cite{6}), seems to dominate over their slight squeeze reduction from the \emph{tangentially} displaced $Cu^{2+}$ neighbors.  Accordingly the combined effect provides some extra squeeze (beyond that in the undoped situation, illustrated near the top or bottom of the frame in the figure) that causes laterally confined Coulomb oscillations of $3s$ electrons through the $O^{2-}$ nuclei in columns 2 and 2' and through the intervening region (across column 1). This is depicted again in Fig. 1 by straight-line hatching.  Third, the opposite scenario holds for the $O^{2-}$ ions of column 3.  Here the squeeze reduction from tangentially more displaced $Cu^{2+}$ neighbors apparently dominates over additional squeeze mutually incurred with nearest-neighbor $O^{2-}$ ions between columns 2 and 3.  The combined effect thus falls short of extra squeeze of the $O^{2-}$ ions in column 3.  Their $3s$ electrons then overswing laterally (not shown in Fig. 1) and diffuse in the $CuO_{2}$ plane, like other $3s$ electrons from $O^{2-}$ ions without extra squeeze, as discussed above.
Graphically, the white squarish areas between columns 2 and 3 imply the corresponding lack of confined Coulomb oscillations. Also, the energy of those less confined $3s$ electrons is \emph{not} extra elevated, preventing anisotropic tunnelling and thus their imaging by STM.\cite{10}

Now compare the present Fig. 1 with Fig. 4be of Ref. 1:  The ion columns 4 and 4' show unidirectional electron density distribution (from axial Coulomb oscillation in the $b$ direction) that furnish the ``sidepieces'' of the ladders visible in the STM pattern.  The sidepieces span along the $4{a_0} \times 4{a_0}$ domain (here, parallel to the $b$ axis).  Their extension along successive domains (Fig. 4ad) provides pathways for superconductivity throughout the $CuO_{2}$ plane of the compound sample.  According to the Coulomb-oscillator model of superconductivity such pathways must be monatomic in width.  Figures 3, 4 and 6 of Ref. 1 confirm that this is the case:  No lateral leakage occurs along close parallel pathways which would spoil superconductivity by electron diffusion.  The $O^{2-}$ ions along the columns 2 and 2', on the other hand, show isolated, diatomic Coulomb oscillators in the sideways $a$-direction through those $O^{2-}$ nuclei and across the intervening region.  They can be recognized as the ``rungs'' of the ladders in the STM pattern.  The rungs don't contribute to superconductivity due to their lack of connectivity.  

The qualitative agreement of the present Fig. 1 with Fig. 4be of Ref. 1 lends support to the validity of the Coulomb-oscillator model of superconductivity.

\bigskip

\centerline{ \textbf{ACKNOWLEDGMENTS}}

\noindent I thank Duane Siemens for stimulating discussions and Preston Jones for help with LaTeX.

\pagebreak

\includegraphics[width=5.7in]{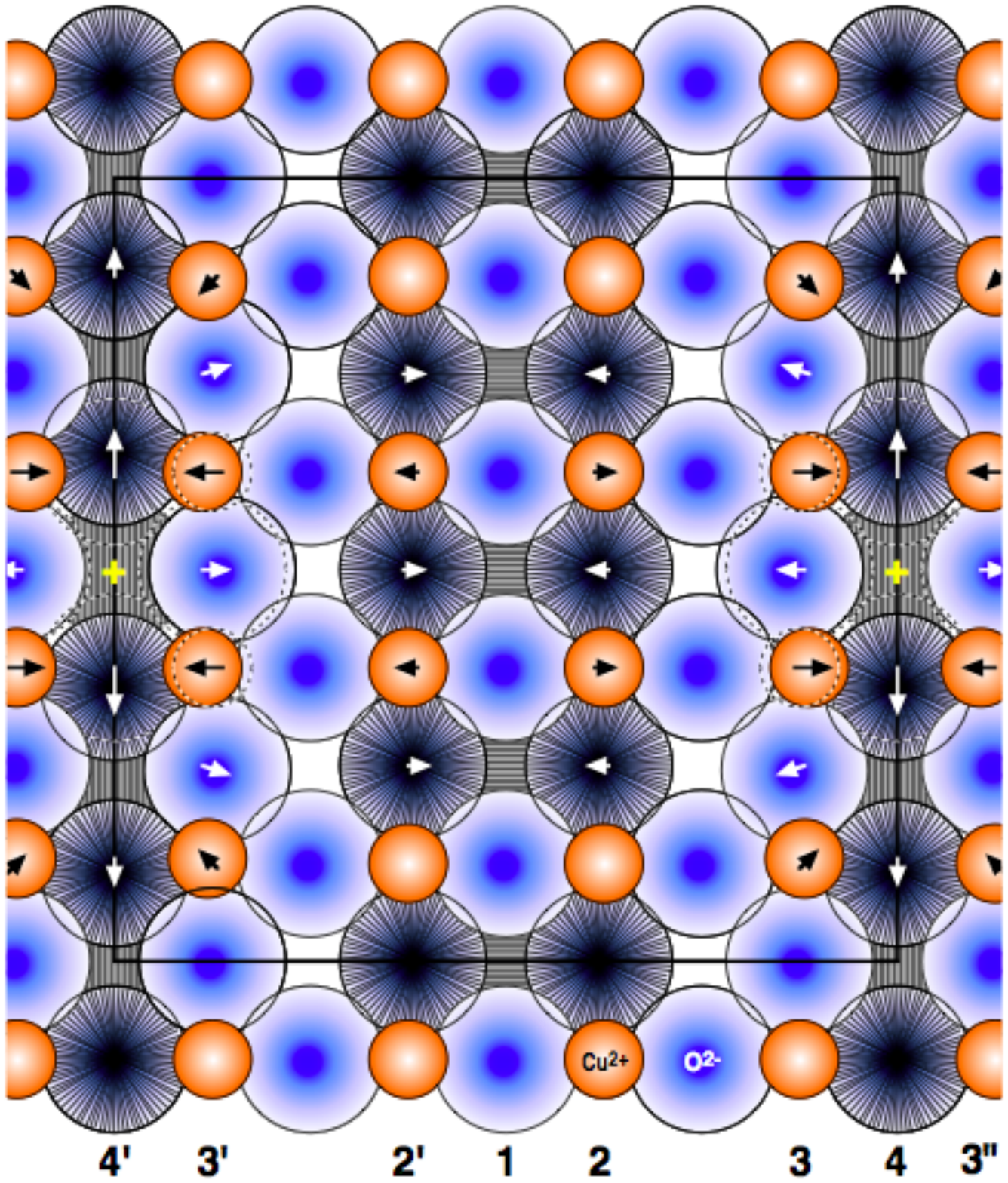}

\noindent FIG. 1. $CuO_{2}$ plane of $Ca_{2-x}Na_{x}CuO_{2}Cl_{2}$, $x = 0.125$, with $4{a_0} \times 4{a_0}$ domain (delimited by the frame).  The crosses halfway up the sides of the frame each mark the position of a doped $Na^{+}$ ion pair residing in the plane above and beneath.  Ions are shown by their hard-sphere size.  Arrows indicate prominent ion displacements.  Dashed circles show positions in the undoped crystal for some ions. Straight-line hatching indicates Coulomb oscillation of $3s$ electrons
through oxygen nuclei and, laterally confined, across the interionic region.


\begin{thebibliography}{12}
\bibitem{8} In order to highlight the relevant mechanisms, only the largest displacements are marked in Fig. 1 by arrows, but not near the top and bottom of the frame.  This permits near the frame's top or bottom a comparison with the ion configuration of an undoped compound.
\bibitem{9} Slightly reducing the large effect from the $Cu^{2+}$ neighbors, the $O^{2-}$ ions of column 4 are, on average, less squeezed by their soft nearest $O^{2-}$ neighbors in columns 3 and 3'' (participating squeeze effect, Ref. 6) as they are in an undoped crystal.
\end{thebibliography}
\end{document}